\documentclass[12pt]{article}

\usepackage{amsmath}
\usepackage{amssymb}
\usepackage{mathrsfs}
\usepackage{graphicx}
\usepackage{enumerate}
\usepackage{cite}
\usepackage{caption}
\usepackage{subcaption}
\usepackage[percent]{overpic}

\def\R{\mathbb{R}}

\newcommand\tS{{\tilde S}}
\newcommand\half{\frac{1}{2}}
\newcommand\eps{\epsilon}
\newcommand\be{\begin{equation}}
\newcommand\ee{\end{equation}}

\usepackage[pdftex,colorlinks=true,linkcolor=blue,citecolor=blue,urlcolor=blue]{hyperref}

\usepackage{fancyhdr}
\fancypagestyle{title}{%
  \setlength{\headheight}{22pt}%
  \fancyhf{}
  \fancyhead[R]{BRX-TH-6314}
}%

\begin{document}

\title{\bf Entanglement shadows in LLM geometries}
\author{Vijay Balasubramanian\footnote{vijay@physics.upenn.edu}$\ {}^{a,d}$, Albion Lawrence\footnote{albion@brandeis.edu}$\ {}^b$, \\ Andrew Rolph\footnote{andrewrolph@brandeis.edu}$\ {}^b$, and Simon F. Ross\footnote{s.f.ross@durham.ac.uk}$\ {}^c$ \\   \\
{\it ${}^a$ David Rittenhouse Laboratories, University of Pennsylvania} \\ {\it 209 S 33rd Street, Philadelphia, PA 19104, USA } \\   \\ 
{\it ${}^b$ Martin Fisher School of Physics,} \\{\it  Brandeis University, Waltham, MA 02453, USA }\\ \\
{\it ${}^c$ Centre for Particle Theory, Department of Mathematical Sciences } \\
{\it Durham University, South Road, Durham DH1 3LE, UK} \\ \\
{\it ${}^d$  Theoretische Natuurkunde, Vrije Universiteit Brussel (VUB),} \\
{\it and International Solvay Institutes, Pleinlaan 2, B-1050 Brussels, Belgium} \\
}

\maketitle
\thispagestyle{title}
\abstract{We find a new example of an asymptotically $AdS_5 \times S^5$ geometry which has an entanglement shadow: that is, a region of spacetime which no Ryu-Takayanagi minimal surface enters. Our example is a particular case of the supersymmetric LLM geometries.  Our results illustrate how minimal surfaces, which holographically geometrize entanglement entropy, can fail to probe the whole of spacetime,  posing a challenge for attempts to directly reconstruct holographic geometries from the entanglement entropies of the dual field theory.
We also comment on the relation to previous investigations of minimal surfaces localised in the $S^5$ factor of AdS$_5 \times S^5$.}  

\clearpage

\section{Introduction}

The Ryu-Takayanagi proposal \cite{Ryu:2006bv}\ and its generalizations provide a map between quantum entanglement of spatial regions of a strongly coupled large-N field theory and the spacetime geometry of its gravitational dual, by relating entanglement entropies to areas of minimal or extremal \cite{Hubeny:2007xt}\ surfaces.  This has led to explicit progress in bulk reconstruction, particularly for linearized perturbations of anti-de Sitter space \cite{Faulkner:2013ica,Czech:2016xec,deBoer:2016pqk}.    There are also attempts to directly represent the areas of arbitrary surfaces in asymptotically AdS spacetimes in terms of new information theoretic observables such as ``differential entropy'' \cite{Balasubramanian:2013lsa,Myers:2014jia,Headrick:2014eia}.   All of these efforts explicitly use the geometry and deformations of extremal surfaces of holographic geometries.

This program is complicated (or enriched, as the reader prefers) by the existence of entanglement shadows: regions of the bulk spacetime which are not reached by any minimal or extremal surfaces used to compute entanglement between spatial regions of the boundary (see, e.g., \cite{Balasubramanian:2014sra,Freivogel:2014lja}).  Holographic geometries with entanglement shadows require additional quantities beyond spatial entanglement in the dual field theory for the purpose of  bulk geometry reconstruction.\footnote{One might likewise wonder how an entanglement-based program would be extended to the BFSS model \cite{Banks:1996vh}, which is a quantum-mechanical model with an 11d holographic dual \cite{Balasubramanian:1997kd,Polchinski:1999br}.}   Consider, for example, the conical defect spacetimes describing excitations of $AdS_3$.   In this case, {\it non-minimal} extremal surfaces enter the entanglement shadow region, and there is a candidate generalization of spatial entanglement called {\it entwinement} which yields quantities dual to the area of these surfaces \cite{Balasubramanian:2014sra,Balasubramanian:2016xho,Lin:2016fqk}.  

In this work, we will argue that a simple but topologically non-trivial asymptotically AdS$_5 \times S^5$ geometry has an entanglement shadow. 
Our example is one of the ``LLM geometries'' \cite{Lin:2004nb}, which are holographically dual to 1/2-BPS excitations of $\mathcal N=4$  super-Yang Mills theory.
These geometries are smooth but topologically complex, and the map to the  dual field theory state is known precisely.    From the perspective of reconstructing bulk geometry from quantities in the dual field theory, one of the most interesting aspects of the LLM geometries is that they are inherently 10-dimensional -- there is no factorization into an asymptotically AdS$_5$ part and a compact part.  If there were such a factorization, we could ``compactify'' the reconstruction problem to one of just recovering the geometry and fields in the asymptotically AdS factor, but that is not possible here.  In fact, it is known that reconstructing the interior geometry and topology of LLM spacetimes  from the dual field theory  using just local operator measurements would require access to trans-Planckian physics \cite{Balasubramanian:2006jt, Balasubramanian:2007zt}.  In particular, around configurations with non-trivial topology there is entanglement between the effective dynamical degrees of freedom and UV modes that are beyond the Planck scale \cite{Berenstein:2017abm, Berenstein:2016mxt, Berenstein:2016pcx}.


We will consider an LLM geometry which is approximately AdS$_5 \times S^5$ in both the asymptotic region and a central region in the spacetime.  In our geometry, the $S^3$ radial sections of the asymptotic AdS$_5$ essentially exchange roles with an $S^3$ factor inside the $S^5$ to form the central AdS$_5$ region.  In this geometry, we study minimal surfaces anchored at the equator of the $S^3$ on the spacetime boundary; these are expected to be the deepest minimal surface probes of the geometry, and compute the entanglement entropy of half the field theory with the other half.
Because of the exchange of the roles of the $S^3$ factors which we described above, a surface that partitions the boundary of AdS in the asymptotic region will partition the $S^5$ in the central region. 
Making some systematic approximations,
 we find that in this central region, the minimal surface for a boundary condition which divides the $S^5$ penetrates into the bulk only for a proper radial distance of order one in the central AdS factor. At this distance, this surface closes off by reaching the pole on the $S^5$. From the point of view of the full LLM geometry, this implies that essentially the whole of the central IR region is not accessed by boundary-anchored minimal surfaces.  This is our shadow region. We then argue that there is an extremal {\it non-minimal} surface, also anchored at the equator of the spacetime boundary, which does enter the shadow region.   This is similar situation as for the conical defects in AdS$_3$ which have an entanglement shadow which is penetrated by non-minimal, but extremal, surfaces \cite{Balasubramanian:2014sra}. 


 Unlike in AdS$_3$ \cite{Balasubramanian:2016xho,Lin:2016fqk} we do not yet have a candidate information theoretic quantity such as {\it entwinement}  that computes the area of a non-minimal  extremal surface from the perspective of the dual field theory.
 The idea in \cite{Balasubramanian:2016xho,Lin:2016fqk}  was that non-minimal extremal surfaces (``long'' geodesics in that case) were related to entanglement in a partition of degrees of freedom of the dual field theory that was not spatially organized.    It would be worth understanding whether there is such an interpretation for non-minimal extremal surfaces in the AdS$_5$ case also.     Interesting earlier holographic studies  of Yang-Mills theory in the Coulomb branch  \cite{Mollabashi:2014qfa,Graham:2014iya,Karch:2014pma} had proposed that  a minimal surface which divides the $S^5$ part of the boundary of an asymptotically AdS$_5 \times S^5$ spacetime can be identified with the entanglement entropy associated to a non-spatial division of the field theory degrees of freedom.  
  In the context of our geometries, the surfaces described by   \cite{Mollabashi:2014qfa,Graham:2014iya,Karch:2014pma} can be regarded as extremal surfaces in the central AdS$_5$ region, which can be extended into the asymptotic AdS$_5$ region to describe entanglement in a conventional spatial partition of the UV theory.  Our analysis shows that the particular extremal surfaces studied in  \cite{Mollabashi:2014qfa,Graham:2014iya,Karch:2014pma}  are not, in fact, the minimal ones that are asymptotic to the equator of the boundary $S^5$.     It would be very interesting to understand
 what information theoretic quantity is being computed by such extremal surfaces, and also by the true minimal surfaces with these boundary conditions.


Our paper is organized as follows.  In section \ref{llm}, we briefly review the LLM geometries first constructed in \cite{Lin:2004nb}, and introduce the examples we consider. In section \ref{irext} we consider extremal surfaces in the central region of our geometries, and explain the relation to the earlier work of \cite{Mollabashi:2014qfa,Graham:2014iya,Karch:2014pma}. In section \ref{shadow}, we argue that the boundary-anchored minimal surfaces in our spacetime close off on the central $S^5$ without penetrating deep into the central region.  Hence  these LLM geometries have entanglement shadows. In section \ref{disc}, we discuss the extension to other LLM geometries and the interpretation of our results.

\section{LLM geometries}
\label{llm}

The 1/2 BPS solutions found by Lin, Lunin and Maldacena (LLM) 
\cite{Lin:2004nb} provide a rich class of asymptotically AdS$_5 \times S^5$ spacetimes where the geometry can be analyzed analytically, and for which precise field theory duals are known. We will focus on a simple example in this class, and find the bulk extremal surface whose area computes the entanglement between halves of the spatial $S^3$, across an equator, in the dual field theory.

The LLM geometries correspond to 1/2 BPS states in $\mathcal N =4$ SYM on $S^3 \times \mathbb R$, where the energy of the state ($\Delta$) is equal to the charge ($J$) under a $U(1)$ subgroup of the $SO(6)$ R-symmetry, $\Delta = J$. The dual geometries should thus be asymptotically AdS$_5 \times S^5$ solutions preserving half the supersymmetry, the $SO(4)$ rotational symmetry on the spatial $S^3$, an $SO(4)$ subgroup of the R-symmetry, and a diagonal $\mathbb R$ group which combines time translation with the $U(1) \in SO(6)$ to leave the state invariant. LLM found that these restrictions fix the form of the geometry up to a single function of three coordinates, $z(y,x_1,x_2)$ \cite{Lin:2004nb}. The metric is     
\begin{equation} \label{LLM}
ds^2 = - h^{-2} (dt+ V_i dx^i)^2 + h^2 (dy^2 + dx_1^2 + dx_2^2) + y e^G d\Omega_3^2 + y e^{-G} d\tilde \Omega_3^2, 
\end{equation}
$i=1,2$, where the functions $h$ and $G$ are related to $z$ by 
\begin{equation}
h^{-2} = 2y \cosh G, \quad z = \frac{1}{2} \tanh G, \label{eq:llmfunctions}
\end{equation}
and $V_i$ is determined by 
\begin{equation}
y \partial_y V_i = \epsilon_{ij} \partial_j z, \quad y(\partial_i V_j - \partial_j V_i) = \epsilon_{ij} \partial_y z.
\end{equation}
The geometry is supported by a self-dual five-form; the explicit form of the field strength is not needed here.
%
Note that in these coordinates the length element $ds^2$ has units of length, as do $y,x_1,x_2$, while $t$ is a dimensionless quantity. 

The range of the coordinates is $y \in (0, \infty)$, $x^i \in (-\infty, \infty)$, so this is an upper half space. The metric and five-form give a solution of the supergravity equations of motion if the function $z$ obeys
\begin{equation} \label{zeq} 
\partial_i \partial_i z + y \partial_y \left( \frac{\partial_y z}{y} \right) = 0. 
\end{equation}
The solutions will be smooth if $z$ satisfies the boundary condition $z \to \pm 1/2$ as $y \to 0$. The general solution of \eqref{zeq} with such boundary conditions was given in \cite{Lin:2004nb}. Hence solutions are specified by giving a colouring of the $x_1, x_2$ plane, specifying regions where $z \to 1/2$, which we will draw in white, and regions where $z \to -1/2$, which we will draw in black. The regions where $z \to -1/2$ correspond to the first $S^3$, with metric $d\Omega_3^2$, shrinking to zero as $y \to 0$, while $z \to 1/2$ corresponds to the second $\tilde S^3$, with metric $ d\tilde \Omega_3^2$, shrinking to zero. 

Note that the solution is a ten-dimensional geometry, and except in special cases, it is not possible to straightforwardly perform a Kaluza-Klein reduction to obtain a five-dimensional description; we really need to think about these geometries using a ten-dimensional perspective. 

\subsection{AdS$_5 \times $S$^5$}

\begin{figure} \label{Disks}
    \centering
    \begin{subfigure}[b]{0.3\textwidth}
        \includegraphics[width=\textwidth]{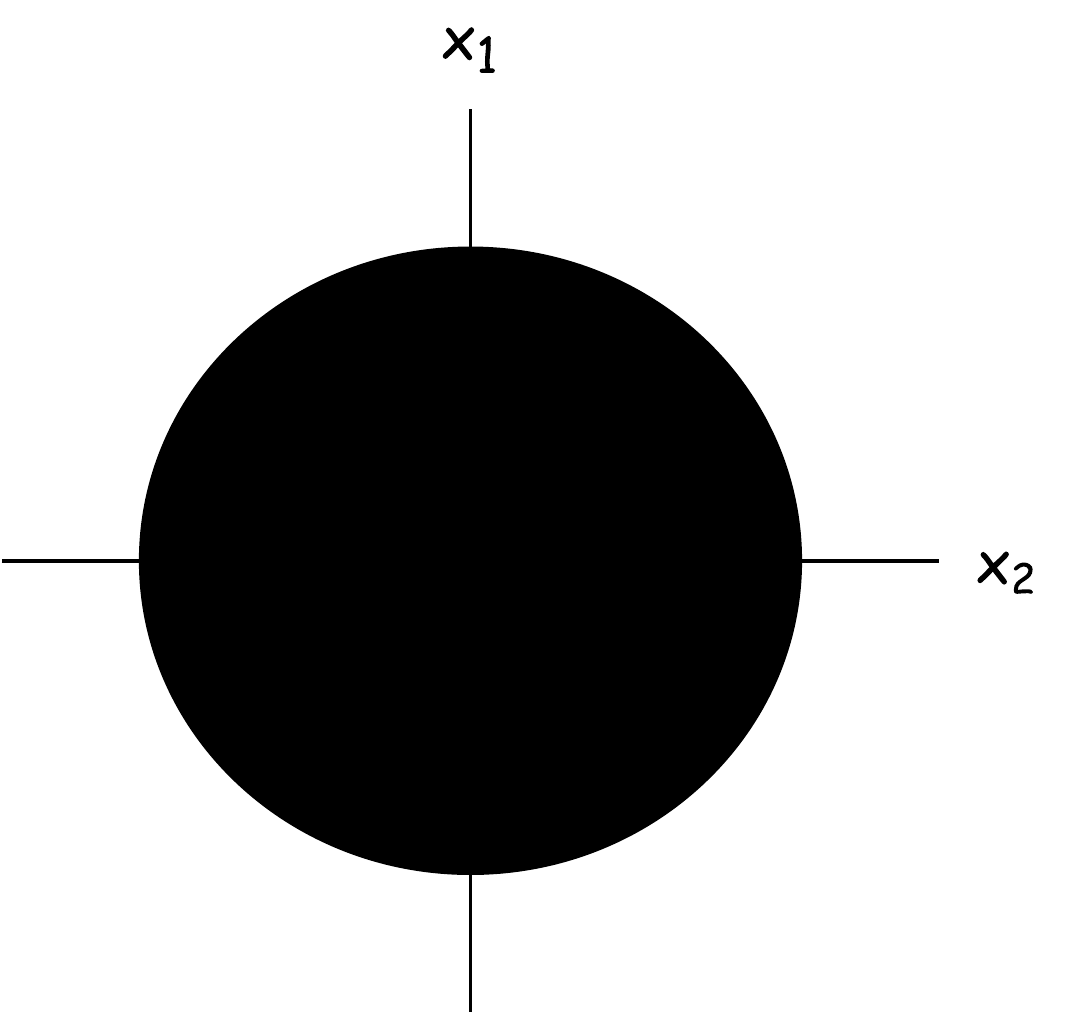}
        \caption{Disk}
        \label{Disk}
    \end{subfigure}
    ~ 
    \begin{subfigure}[b]{0.3\textwidth}
        \includegraphics[width=\textwidth]{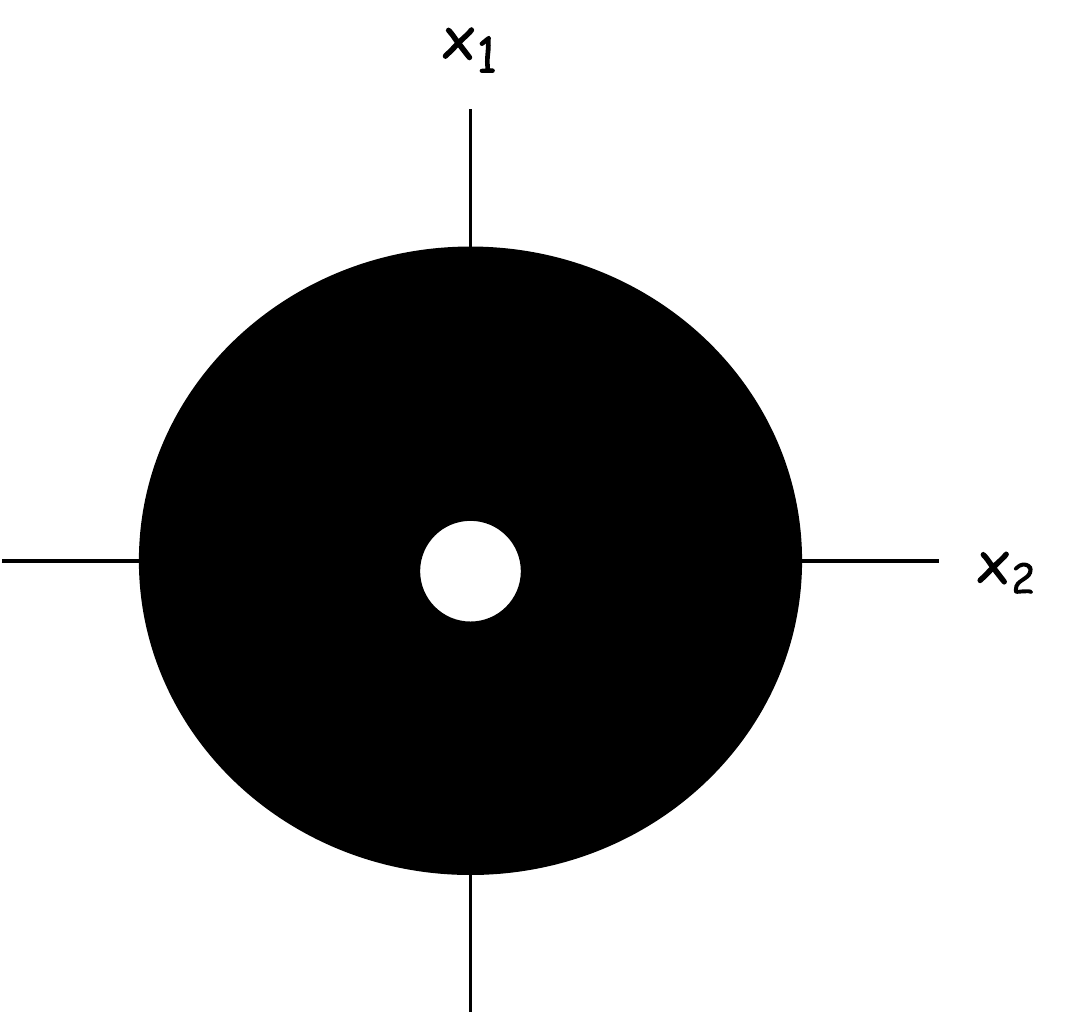}
        \caption{Annulus}
        \label{Annulus}
    \end{subfigure}
    ~ \caption{LLM configurations in the $(x_1,x_2)$ plane. The configurations describe boundary conditions for the equations of motion on a two dimensional surface in the bulk spacetime,
and also correspond to configurations in a fermionic phase space that  completely summarizes the  boundary $1/2$ BPS state. The black disc boundary condition (a) leads to a pure $AdS_5 \times S^5$ geometry. We will show that no entangling surface can probe deeply into the IR region of the geometry given by the annulus boundary condition (b).}
\end{figure}

The simplest example is the disc, where $z= -1/2$ for $r < R$, and $z=1 /2$ for $r > R$, where $r^2 = x_1^2 + x_2^2$. The configuration is shown in Figure~\ref{Disk}. The solution for $z$ is  \cite{Lin:2004nb}
\begin{equation}
z = \frac{ r^2 + y^2 - R^2}{2\sqrt{(r^2 + y^2 + R^2)^2 - 4r^2 R^2}}. 
\end{equation}
This corresponds to the vacuum AdS$_5 \times S^5$ solution. If we make the change of coordinates
\begin{equation} \label{ry}
y = R \sinh \tilde \chi \sin \tilde \theta, \quad r = R \cosh \tilde \chi  \cos \tilde \theta,  \quad \tilde \phi = \phi -t, 
\end{equation}
where $\phi$ is the angular coordinate in the $x_1, x_2$ plane, the metric becomes: 
\begin{equation} \label{uvads}
ds^2 = R( - \cosh^2 \tilde \chi dt^2 + d \tilde \chi^2 + \sinh^2 \tilde \chi d\Omega_3^2 +  d\tilde \theta^2 + \cos^2  \tilde \theta d\tilde \phi^2 + \sin^2 \tilde \theta d\tilde \Omega_3^2).
\end{equation}
The first three terms describe the metric on $AdS_5$ with $AdS$ radius $R$; the last three terms describe the metric on $S^5$ with constant radius $R$.

In these coordinates, for $r < R$, $y = 0$ corresponds to $\tilde \chi = 0$, while for $r > R$, $y = 0$ corresponds to $\tilde \theta = 0$. Thus, the black disc $r <R$ corresponds to the origin in the AdS factor, with position on the disc mapping to position on the $S^5$. The fibration of ${\tilde S}^3$ over a hemisphere surrounding this disc is topologically an $S^5$, homologous to the $S^5$ factor in the geometry (\ref{uvads}).

\begin{figure}
\centering 
\includegraphics[width=8cm, trim={0.6cm 2.8cm 2cm 3.5cm},clip]{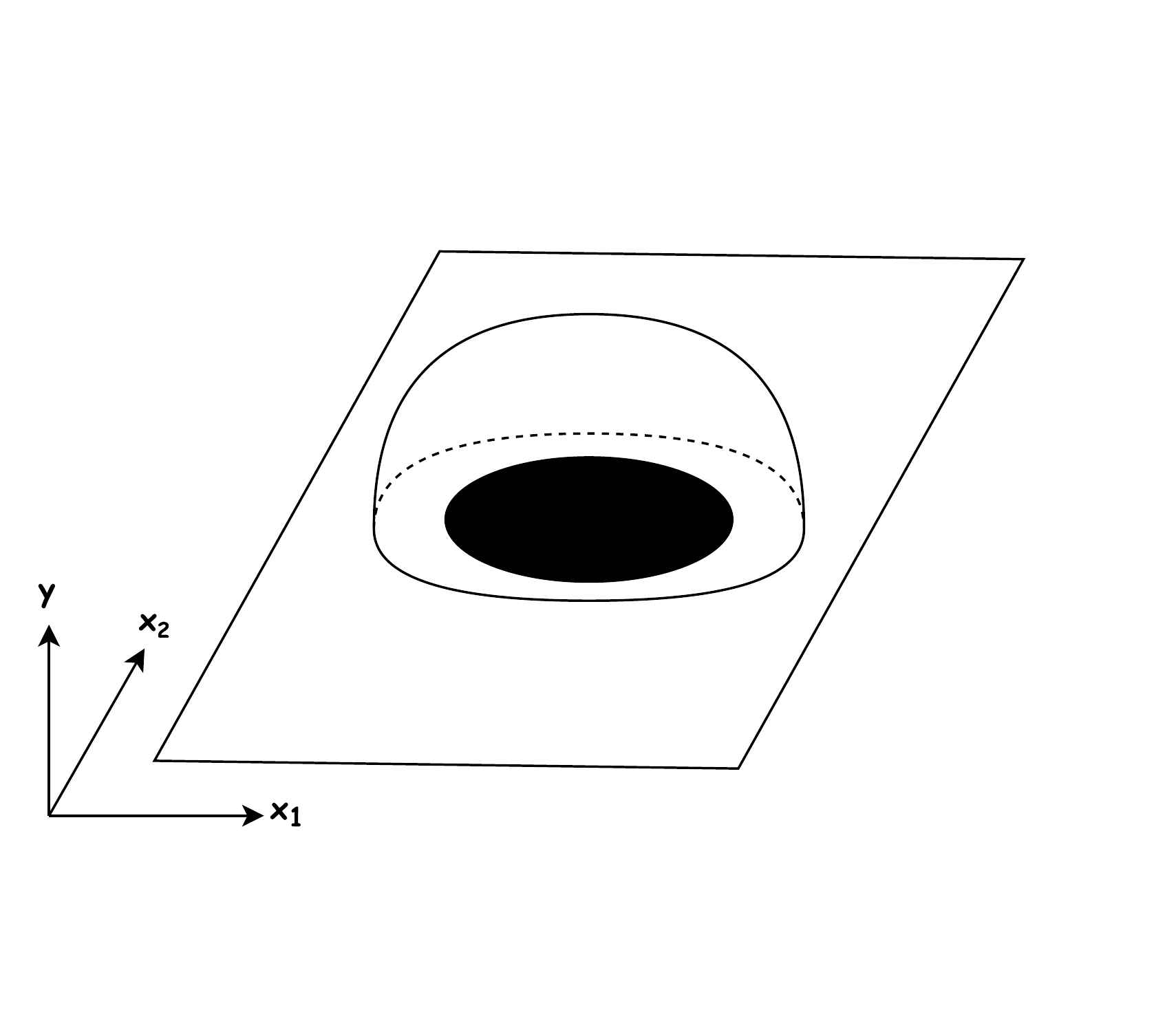}
\caption{A hemisphere over a black disc in the $(x_1,x_2)$ plane. An ${\tilde S}^3$ fibration over this surface is topologically an $S^5$.
}
 \label{S3_fibration}
\end{figure}

A partial visualization is shown in Figure~\ref{S3_fibration}. We can invert the coordinate transformation to write the AdS radial coordinate in general as 
\begin{equation}
\sinh^2 \tilde \chi = \frac{1}{2R^2} (y^2 + r^2  - R^2 + \sqrt{ (y^2 + r^2 + R^2)^2 - 4 r^2 R^2}) .
\end{equation}
At $y^2 + r^2 \gg R^2$, $\sinh^2 \tilde \chi \approx (y^2 + r^2)/R^2$, so the round hemispheres shown for large $r^2 + y^2$ are approximately surfaces of constant radius in the AdS$_5$ factor, but at $y^2 + r^2 \ll R^2$, $\sinh^2 \tilde \chi \approx y^2/R^2$, so the planes of constant $y$ are approximately constant AdS radius, approaching $\tilde \chi =0$ in the black disc.

\subsection{Annulus} 

Perhaps the simplest nontrivial LLM geometry, and the one we will consider, is described by an annulus in the $x_1-x_2$ plane, with a white disc inside the black one (Figure~\ref{Annulus}). That is, we take the boundary conditions for the function $z$ to be $z(y=0,r>R) = \frac{1}{2}$, $z(y=0,R > r > \epsilon) = -\frac{1}{2}$, and $z(y=0,r< \epsilon) = \frac{1}{2}$. The solution is then 
\begin{eqnarray} 
z &=& \frac{1}{2} - \frac{y^2}{\pi} \int_\epsilon^R \frac{r' dr' d\phi'}{[r^2 + r'^2 -2r r' \cos \phi' + y^2]^2} \\ &=& \frac{1}{2} + \frac{1}{2} \frac{r^2 +y^2 - R^2}{\sqrt{(r^2 + y^2 + R^2)^2 - 4 r^2 R^2}} - \frac{1}{2} \frac{r^2 +y^2 - \epsilon^2}{\sqrt{(r^2 + y^2 + \epsilon^2)^2 - 4 r^2 \epsilon^2}}. \label{ann}
\end{eqnarray}

The physical picture of this configuration is that it represents the back-reacted version of maximal giant gravitons \cite{McGreevy:2000cw}.  We consider a set of D3-branes wrapping the $\tilde S^3$ inside $S^5$ at $\tilde \theta =\pi/2$ where this $\tilde S^3$ has maximal volume, with angular momentum along $\tilde \phi$ corresponding to the R-charge.    These D-branes dissolve into the backreacted geometry.  Note that because of the angular momentum, the annulus geometry is stationary but not static.  

\subsection{Approximating the annulus geometry}

Extremal surfaces in the annulus geometry are in general complicated, and finding them involves solving a nonlinear PDE in two variables.  To make the problem more tractable, we will consider the case where $\epsilon \ll R$, so that the white hole in the center of Fig.~\ref{Annulus} is small compared to the area of the outer disk.  We will still consider the case that both radii are large compared to the string or Planck scales. The result is a separation of scales that leads to a straightforward picture of the geometry.

To begin with, we can consider the coordinates at ``large" radius, for which  $r^2 + y^2 \gg \epsilon R$. In this case, the white disk in the center will appear small and we expect the geometry to be a small perturbation of an $AdS_5\times S^5$ geometry with radius of curvature $R$.  More precisely, the last term in \eqref{ann} can be approximated by a series expansion in $\epsilon^2/(r^2+y^2)$, which at leading order gives 
\begin{equation} \label{extAdS}
z \approx \frac{1}{2} \frac{r^2 +y^2 - R^2}{\sqrt{(r^2 + y^2 + R^2)^2 - 4 r^2 R^2}} + \frac{\epsilon^2 y^2}{(r^2 + y^2)^2}. 
\end{equation}
The corresponding geometry is the $AdS_5\times S^5$ metric \eqref{uvads} with a subleading correction which decays at large distances. We call this the ``UV $AdS$ region".

On the other hand, if we consider small distances $r^2+y^2 \ll \epsilon R$, the geometry is well approximated by a black plane with a white disk in the center.  Now the LLM geometries are symmetric under $z = \half \to - \half$ while exchanging $S^3$ and 
$\tS^3$; thus, the region is well approximated by $AdS_5\times S^5$ with radius of curvature $\epsilon$, which we dub the ``IR AdS region". More precisely, the second term in \eqref{ann} can be expanded in a series in $1/R^2$, which gives 
\begin{equation} \label{closetoholez}
z \approx - \frac{1}{2} \frac{r^2 +y^2 - \epsilon^2}{\sqrt{(r^2 + y^2 + \epsilon^2)^2 - 4 r^2 \epsilon^2}} + \frac{y^2}{R^2}.
\end{equation}
The final term in \eqref{closetoholez} gives a correction to IR AdS which decays in the interior, and grows as we move to large distances. The fact that the sign of the leading term in $z$ is reversed as compared to (\ref{extAdS}) means that $\tilde S^3$ is now the sphere factor in the IR AdS space, while $S^3$ is the sphere factor in the $S^5$. If we further adopt the AdS coordinates
\begin{equation} \label{IRcoords}
y = \epsilon \sinh \chi \sin \Theta, \quad r = \epsilon \cosh \chi \cos \Theta, \quad \tilde \phi = \phi + t, 
\end{equation}
the leading order metric is 
\begin{equation} \label{irads}
ds^2 = \epsilon ( - \cosh^2 \chi dt^2 + d\chi^2 + \sinh^2 \chi d\tilde \Omega_3^2 + \cos^2 \Theta d\tilde \phi^2 + d\Theta^2 + \sin^2 \Theta d\Omega_3^2
)\ ,
\end{equation}
making the approximate $AdS_5\times S^5$ geometry explicit.

The IR AdS geometry can be thought of as the back-reacted description of the D3-branes in the giant graviton picture mentioned in \S2.2 above. The D3-branes wrap the $\tilde S^3$, so this becomes the spatial directions in the AdS factor in this IR geometry. 

One might hope that these two descriptions have an overlapping regime of validity, where $\epsilon^2 \ll r^2 + y^2 \ll R^2$. However, in this intermediate regime, 
\begin{equation}
z \approx -\frac{1}{2} + \frac{y^2}{R^2} + \frac{\epsilon^2 y^2}{(y^2 + r^2)^2}\ .\label{eq:intz}
\end{equation}
The second and third terms are small, but as features of the geometry depend on $z + \frac{1}{2}$, we cannot neglect either of them. We therefore need to analyze the behavior in this region independently. As an indication, consider the volume of the spheres $S^3,\tS^3$. The $S^3$ volume is:
\begin{equation}
y e^G \approx \frac{y^2}{R} \sqrt{ 1 + \frac{\epsilon^2 R^2}{(y^2 + r^2)^2}},  
\end{equation}
and the $\tilde S^3$ volume is 
\begin{equation}
y e^{-G} \approx R \left( 1 + \frac{\epsilon^2 R^2}{(y^2 + r^2)^2} \right)^{-1/2}. 
\end{equation}
To arrive at these we solved for $e^G$ using (\ref{eq:llmfunctions}) and the approximation (\ref{eq:intz}) for $z$, throwing out terms that are higher order in $\eps/R$.
If $y^2 + r^2 \sim \epsilon R$, the terms inside the square roots in each equation are all of order ${\cal O}(1)$, and the square roots cannot be approximated as constants. Thus, this region is not well covered by either the UV or the IR AdS approximation. 

Instead, a natural coordinate system in this region is:
\begin{equation}
y = \sqrt{\epsilon R} e^\zeta \sin \Theta, \quad r = \sqrt{\epsilon R} e^\zeta \cos \Theta, 
\end{equation}
so that $y^2 + r^2 \sim \epsilon R$ corresponds to $\zeta$ near zero. This is essentially a rescaled version of the IR coordinates \eqref{IRcoords}, with $e^\chi = \sqrt{\frac{R}{\epsilon}} e^\zeta$. Then 
\begin{equation}
y e^G \approx \epsilon e^{2\zeta} \sin^2 \Theta \sqrt{1 + e^{-4 \zeta}}, 
\end{equation}
and the $\tilde S^3$ volume is 
\begin{equation}
y e^{-G} \approx R \left( 1 + e^{-4\zeta}  \right)^{-1/2}. 
\end{equation}
The function
\begin{equation}
h^{-2} = y e^G + y e^{-G} \approx y e^{-G} = R \left( 1 + e^{-4\zeta}  \right)^{-1/2}.
\end{equation}
Using (\ref{eq:intz}), we find
\be
	V_{\phi} = \frac{\eps^2 r^2}{(r^2 +y^2)^2}
\ee
which is order ${\cal O}\left(\frac{\eps}{R}\right)$.  If we further rescale $t \to \sqrt{\frac{\eps}{R}} t$, then the $g_{t\phi}$ terms are of order $\eps\sqrt{\frac{\eps}{R}}$, and the term $V_{\phi}^2 d\phi^2$ is of order $\eps\frac{\eps}{R}$; these can be neglected as the remaining terms are of order $\eps$.  

Thus, using $dy^2 + dr^2 = \epsilon R e^{2 \zeta} (d \zeta^2 + d\Theta^2)$, the metric is to order ${\cal O(\eps)}$,
\begin{eqnarray} 
& & ds^2 \approx - \frac{\eps}{\sqrt{1 + e^{-4\zeta}}} dt^2 \nonumber\\
& & \qquad \qquad +  \epsilon \sqrt{1+ e^{-4 \zeta}} e^{2\zeta} ( d\zeta^2 + d\Theta^2 + \cos^2 \Theta d\phi^2 + \sin^2 \Theta d\Omega_3^2) \nonumber\\
& & \qquad \qquad + R (1 + e^{-4 \zeta})^{-1/2} d\tilde \Omega_3^2. \label{int} 
\end{eqnarray}
This metric is static and stationary up to corrections that are down by powers of $\sqrt{\frac{\eps}{R}}$.
The approximations leading to this form of the metric hold if $\eps^2 \ll y^2 + r^2 \ll R^2$, so that we can up to a point take $\zeta \ll 0$. In this limit, we regain the large $\chi$ part of the IR metric \eqref{irads} plus small corrections. Note the particular simplification in this intermediate region: the coordinates of the the IR $S^5$ are multiplied by the same radial factor, so that the geometry still has the $SO(6) \times SO(4)$ symmetry of the IR region.  

Thus, we have {\it three} approximate descriptions: the IR AdS description, \eqref{irads}, valid for $r^2 + y^2 \ll \epsilon R$, the intermediate description \eqref{int}, valid for $\epsilon^2 \ll r^2 + y^2 \ll R^2$, and the UV AdS description \eqref{uvads}, valid for $r^2 + y^2 \gg \epsilon R$.     Between the UV and IR AdS descriptions there is an exchange  of spheres:  the $S^3$ in the asymptotic AdS factor exchanges roles in the IR geometry with an $S^3$ that is  the $S^5$ factor of the asymptotic geometry.  We will use these three overlapping descriptions to analyze the minimal surfaces. 

\section{Extremal surfaces in empty $AdS_5\times S^5$}
\label{irext}

As we have discussed, the annulus geometry for $\eps \ll R$ interpolates between two $AdS_5\times S^5$ regions, a ``UV" region with radius of curvature $R$, and an ``IR" region, with radius of curvature $\eps$.  In this background, we are interested in finding extremal surfaces which are anchored at  the equator of the UV boundary, bisecting the $S^3$ of the asymptotically AdS factor of the geometry.  Because of the symmetry of the problem,  and taking $\theta$ to be polar angle on this $S^3$ ($\Omega_3$ in the metric (\ref{uvads})), there is an extremal surface at fixed $t = t_0, \theta = \frac{\pi}{2}$ which extends from the UV region into the IR region.  In the UV region this surface wraps the $S^5$ factor and bisects the $S^3$ of the asymptotic AdS$_5$.  As we discussed above, in the IR region, the $S^3$ of the  asymptotic AdS$_5$ exchanges roles with an $S^3$ inside the $S^5$.  Thus, in the IR region (\ref{irads}) a fixed $\theta = \frac{\pi}{2}$ surface wraps the $\tS^3$ of the AdS factor, while bisecting the $S^5$ factor.

%

As we will show, this is not the actual minimal surface for the LLM geometry.   To understand why, it will be helpful to first consider the co-dimension two spacelike minimal surfaces of empty AdS$_5 \times S^5$ with boundary conditions that either bisect the AdS$_5$ or the $S^5$.    Extremal surfaces bisecting the $S^5$ of AdS$_5 \times S^5$ were previously studied in \cite{Mollabashi:2014qfa,Graham:2014iya,Karch:2014pma} the authors of which were interested in studying non-spatially organized entanglement in the Coulomb branch of gauge theories.  We are interested in such surfaces because in our LLM setting the obvious candidate minimal surface bisects the $S^5$ of the  AdS$_5 \times S^5$ in the interior of the geometry (the ``IR").   
We will show that surfaces that occupy a fixed angular position on the $S^5$ cannot in fact be a minimal surface;  in fact, if we cut off the AdS$_5$ factor by any amount, a minimal surface that partitions the $S^5$ at the cutoff slips off the  sphere over radial distances of order the cutoff.   In our LLM case, this will imply that the minimal surfaces of interest to us, which bisect the asymptotic AdS$_5$ boundary, will slip off the $S^5$ in the deep interior part of the geometry and thus terminate smoothly before penetrating this region.


\subsection{Minimal surfaces bisecting $AdS_5$}

In pure AdS$_5 \times S^5$  the minimal surface that bisects the boundary of the AdS$_5$ factor penetrates all the way to origin of the spacetime; hence there is no entanglement shadow.   Because the geometry is factorized we can see this from just the AdS$_5$ part of the full geometry in (\ref{uvads}).
Let us choose coordinates for the $S^3$ part of AdS$_5$ in (\ref{uvads})  so that the metric on this sphere is
\begin{equation}
d\Omega_3^2 = d\theta^2 + \sin^2 \theta d\Omega_2^2 \, .
\end{equation}
We want to find minimal spacelike co-dimension 2 surfaces in AdS$_5$ which bound the region $\theta \leq \theta_0$ at the boundary ($\tilde\chi \to \infty$ in  (\ref{uvads})); following Ryu and Takayanagi such a surface should compute the entanglement entropy of the region $\theta \leq \theta_0$ in the field theory dual to the space.
We can take the minimal surface to lie $t=0$, and specify it by a function $\theta(\tilde\chi)$ with the boundary condition $\theta \to \theta_0$ as $\tilde\chi \to \infty$. In the LLM coordinates (see the coordinate transformation (\ref{ry})), the minimal surface is thus specified by $\theta(r,y)$ with the boundary condition $\theta \to \theta_0$ as $r^2 + y^2 \to \infty$. Note that the function $\theta$ is typically not defined for all $r,y$; the RT surface will end where $\theta = 0$, that is where it reaches the north pole on the $S^3$. As we increase $\theta_0$, the minimal surface will probe deeper and deeper into the bulk, and the minimal surface for $\theta_0 = \pi/2$, where we keep half of the boundary, should be simply $\theta = \pi/2$ everywhere, slicing the AdS factor in half. 

It is obvious by symmetry that $\theta = \pi/2$ is an {\it extremal} surface for the boundary conditions $\theta_0 = \pi/2$, but it is not immediately obvious in these coordinates that it has minimal area. This will be an important distinction later, so we note here that we can make the minimality of the $\theta = \pi/2$ surface manifest via the coordinate transformation
\begin{equation} \label{hypc}
\sinh \rho = \sinh \tilde \chi \cos \theta, \quad \tanh \gamma = \tanh \tilde \chi \sin \theta,
\end{equation}
In these coordinates the AdS$_5$ part of the  metric in   (\ref{uvads}) is  
\begin{equation}
ds^2 = - \cosh^2 \rho \cosh^2 \gamma dt^2 +  d\rho^2 + \cosh^2 \rho (d\gamma^2 + \sinh^2 \gamma  d\Omega_2^2). 
\end{equation}
The extremal surface at $t=0, \theta = \pi/2$ becomes the hyperbolic disc at $\rho = 0$ in these coordinates.  To see that this surface is in fact minimal we start at the boundary of this disc ($\gamma \to \infty$) and observe that if $\theta$ were to change from $\pi/2$ as $\gamma$ decreases, the $\cosh^2\rho$ factor (which is 1 when $\theta = \pi/2$) would increase, and with it the area of the surface.

In this AdS example, we can work with a five-dimensional description, but in general LLM geometries we need to work in a ten-dimensional geometry. The extension of the Ryu-Takayanagi prescription to this ten-dimensional setting is to consider a codimension two spacelike surface in the full ten-dimensional geometry, the area of which is calculated in units of the ten-dimensional Newton's constant \cite{Nishioka:2006gr} (see \cite{Jones:2016iwx} for a fuller discussion). In the present case, the minimal surface in the ten-dimensional description is simply the eight dimensional surface at $t=0$, $\theta = \pi/2$, wrapping the $S^2$ in the $S^3$,  the $S^5$, and the extending along the radial $\tilde \chi$ direction in the AdS factor. In the LLM coordinates (\ref{LLM}), this surface wraps the equatorial $S^2 \subset S^3$ and the entire $\tilde S^3$; and it fills the $y, x_1, x_2$ space.  

\subsection{Extremal surfaces bisecting $S^5$}

Ref. \cite{Mollabashi:2014qfa}\ considered surfaces in AdS$_5 \times S^5$ which slice the $S^5$ in half, while wrapping the AdS$_5$ factor. The authors proposed that such surfaces could be interpreted as geometrizing entanglement between different CFT components on the $S^5$, corresponding to some non-spatial decomposition of the CFT Hilbert space. This was further investigated in \cite{Graham:2014iya,Karch:2014pma}, where it was proposed that it could correspond to decomposing the CFT Hilbert space in terms of R symmetry representations. 
The analysis in  \cite{Mollabashi:2014qfa,Karch:2014pma} was mainly based on going onto the Coulomb branch of the CFT on $\R^4$, where one could define a division of the CFT Hilbert space at low energies into two factors associated with the unbroken gauge group at low energies. But the relationship of entanglement between these factors and geometrical surfaces in the bulk remains conjectural. 

On the other hand, the extremal surfaces that appear in  \cite{Mollabashi:2014qfa,Graham:2014iya,Karch:2014pma} are directly relevant to the holographic representation of spatial entanglement in the annular LLM geometry that we are studying here.
 If we start in the UV region with a spatial decomposition of the field theory along an equator $\theta_0 = \pi/2$, the symmetries of the theory including $t \to - t$ imply that the surface $t = {\rm constant}$, $\theta = \pi/2$ is an extremal surface. In the IR region, there is an effective description in terms of a new CFT dual to the IR AdS geometry. In this IR geometry, as discussed above, the extension of a surface that is asymptotically  at $\theta_0 = \pi/2$ bisects an equator on the $S^5$.   Thus, a surface that bisects the $S^5$ of the IR AdS space becomes related to a surface that bisects the AdS$_5$ of the UV region and hence to a spatial decomposition of the UV theory.
 Thus, if we understand the details of how the IR CFT embeds in the UV CFT, we might be able to interpret surfaces of the kind studied in \cite{Mollabashi:2014qfa,Karch:2014pma}.
 We will not pursue such a CFT understanding here.  Instead, we will show that in an $AdS_5 \times S^5$ geometry, the minimal fixed-$t$ codimension two surface which bisects the equator of $S^5$ at the boundary of $AdS_5$ does not extend into the interior of the $AdS_5$ factor.  Rather, for any radial cutoff, the surface extends inwards only by an amount of order that cutoff.

Consider $AdS_5 \times S^5$ spacetime with metric (\ref{irads}) (i.e. in  the coordinates on AdS$_5 \times S^5$ that arise in the IR part of the annular LLM geometry).   We can show that in these coordinates $\theta = \pi/2$ corresponds to an equator on the $S^5$, by introducing new coordinates (in analogy to the change to hyperbolic slicing in AdS in \eqref{hypc}).  To this end, set
\begin{equation} \label{s5coord}
\cos \theta' = \sin \Theta \cos \theta, \quad \sin \theta' \cos \beta = \cos \Theta, 
\end{equation}
so that the metric becomes 
\begin{equation}
ds^2 = \epsilon ( - \cosh^2 \chi dt^2 + d\chi^2 + \sinh^2 \chi d{\tilde \Omega}_3^2 + d\theta'^2 + \sin^2 \theta' (d\beta^2 + \cos^2 \beta d \tilde \phi^2 + \sin^2 \beta  d\Omega_2^2 )).  
\label{foursphere}
\end{equation}
The surface at $\theta = \pi/2$ is at $\theta'=\pi/2$, and is nicely exhibited as an equatorial $S^4$ in the $S^5$ in these coordinates (i.e., the metric $d\beta^2 \cdots$ within the final parenthesis in (\ref{foursphere})). 

In this geometry, this surface is not minimal. This is easily seen by considering a surface where $\theta'$ is some function of $\chi$: the induced metric on the surface is 
\begin{equation}
ds^2 = \epsilon ( (1 + (\partial_\chi \theta')^2) d\chi^2 + \sinh^2 \chi d{\tilde \Omega}_3^2 + \sin^2 \theta'(\chi)d\Omega_4^2 ),  
\end{equation}
so the area functional is 
\begin{equation} \label{areaS}
A = \eps^4 V_{S^4} V_{S^3} \int d\chi \sinh^3 \chi \sqrt{ 1 + (\partial_\chi \theta')^2} \sin^4 \theta'. 
\end{equation}
We can choose a non-trivial function $\theta'(\chi)$ satisfying the boundary condition $\theta' \to \pi/2$ as $\chi \to \infty$ which will lower the area; we just need $\sqrt{ 1 + (\partial_\chi \theta')^2} \sin^4 \theta' < 1$. For $\theta' = \pi/2 - \alpha$ for small $\alpha$, this is $\frac{1}{2} (\partial_\chi \alpha)^2 - 2 \alpha^2 <0$. An example of a function of compact support which makes $\Delta A <0$ is $\alpha = \alpha_0 (e^{-\chi + 3} -1)$ for $\chi <3$, $\alpha = 0$ for $\chi >3$.  Thus we know there are other surfaces with smaller area, before even constructing the minimal surface.

To find the form of the true minimal surface, we turn to the Euler-Lagrange equation for the action \eqref{areaS}:
\begin{equation} \label{eltheta}
\partial_{\chi}^2 \theta'  - (1+ (\partial_{\chi} \theta')^2 ) (-3\coth \chi \partial_\chi \theta' + 4 \cot \theta') = 0\ ,
\end{equation}
or in terms of $\chi(\theta')$,
\begin{equation} \label{elchi}
\partial_{\theta'}^2 \chi  + (1+ (\partial_{\theta'} \chi)^2 ) (-3\coth \chi  + 4 \cot \theta' \partial_{\theta' }\chi) = 0\ .
\end{equation}
There are two possibilities for a smooth surface; either the surface extends to $\chi=0$, with some limiting value $\theta' (0) = \theta_{min}$, and $\partial_\chi \theta' = 0$, or it ends by pinching off at the north pole $\theta'=0$ at some $\chi = \chi_{min}$, with $\partial_\chi \theta' \to \infty$ at $\chi_{min}$. It is convenient to analyse the second possibility in terms of $\chi(\theta')$ rather than $\theta'(\chi)$, so that the smoothness condition becomes $\partial_\theta' \chi(0) = 0$. Note that if $\partial_{\chi} \theta'=0$, $\partial_{\chi}^2 \theta' \geq 0$, so a smooth function satisfying this equation can have a maximum but no minima. In particular, if $\theta'=0$ at some $\chi_{min}$ it must be monotonic. 

We expect the true minimal surface to pinch off by reaching $\theta' = 0$. The reason $\theta=\pi/2$ with bisects the $S^5$ is not minimal in this geometry (unlike the surface which bisects the AdS$^5$ factor) is that the size of the $S^3 \subset S^5$ is independent of the radial $\chi$ direction. Thus, instead of extending down to the origin in the AdS factor, the surface can reduce its area by pinching off on the sphere. Since there is no scale to determine the value $\chi_{min}$ at which the pinch-off occurs, it will be determined by the radius at which we cut off the AdS factor.   That is, we expect that over the range between a radial AdS cutoff at some $\chi = \chi_{max}$ and $\chi = \chi_{max} - \Delta \chi$ the minimal surface should move from $\theta \approx \pi/2$ to end at $\theta =0$, where $\Delta \chi$ is independent of $\chi_{max}$. 

Since $\chi$ is expected to remain large over the full range of $\theta'$, we can approximate \eqref{elchi} by
\begin{equation} \label{chieq}
\partial_{\theta'}^2 \chi  + (1+ (\partial_{\theta'} \chi)^2 ) (-3  + 4 \cot \theta' \partial_{\theta' }\chi) = 0. 
\end{equation}

This is independent of $\chi$, which reflects an invariance of the area functional \eqref{areaS} in this approximation under $\chi \to \chi + a$. This is thus a first order equation for $\partial_{\theta'} \chi(\theta')$, which we should solve with the boundary condition $\partial_{\theta'} \chi(0) = 0$.  We can make this equation look nicer by writing $\partial_{\theta'} \chi(\theta') = \tan \alpha(\theta')$; then 
\begin{equation}
\partial_{\theta'} \alpha = 3  - 4 \frac{\tan \alpha}{\tan \theta'} . 
\end{equation}
%
At $\theta' \approx \pi/2$, we can linearize around $\pi/2$ to obtain 
\begin{equation} \label{eq:IR_oscillations}
\theta' = \frac{\pi}{2} - e^{-3\chi/2} (a_1 \cos(\frac{\sqrt{7}}{2} \chi) + a_2 \sin(\frac{\sqrt{7}}{2} \chi)),
\end{equation}
so the solution approaches $\pi/2$ exponentially; if we want $\theta' = \pi/2 - \delta$ at the cutoff $\chi = \chi_{max}$, it will extend to a range of order $\Delta \chi \sim -\frac{2}{3} \ln \delta$.


In the limit $\chi_{max} \to \infty$ with fixed $\delta$, this surface has infinitely less area than the one at $\theta'= \pi/2$. 

\section{Shadow region in annulus LLM geometry}
\label{shadow}

We can now address our main question. Consider the entanglement across the equator of the $S^3$ for super-Yang-Mills on $S^3\times \R$, in the state dual to the annulus LLM geometry with 
$\eps \ll R$ where $\eps$ is much bigger than the Planck and string lengths. What is the configuration of the extremal surface computing this entanglement, and how far into the interior does the surface extend?

Since this geometry is not static (though it is stationary), in principle we need to consider extremal surfaces following the prescription  \cite{Hubeny:2007xt}.  However, in the UV and intermediate regimes, the metric is static up to corrections of multiplicative order $\sqrt{\eps/R}$. If we approximate the metric as static, we will find that the resulting minimal surface at $t= 0$, computed following \cite{Ryu:2006bv}, does not go beyond the intermediate regime into the interior regime.   Thus we can self-consistently use the prescription in \cite{Ryu:2006bv}.  Note also that the surface at fixed $t$ also respects the $t \to -t$, $\phi \to -\phi$ symmetry of the metric, so it is at least an extremal surface.

We thus consider a minimal surface dividing the $S^3$ in (\ref{LLM}) (which is the $S^3$ of the asymptotic AdS$_5$ factor), and wrapping the $\tilde S^3$ (which is inside the asymptotic $S^5$ factor), and filling the $y, x_1, x_2$ space at fixed $t$, possibly up to some terminal 2d hypersurface in that space where the minimal surface closes off on the $S^3$. The geometry preserves an extra $U(1)$ symmetry, because we have not broken the rotational symmetry in the $x_1-x_2$ plane. The minimal surface will then be specified by some function $\theta(y,r)$ giving the polar angle of the surface on the $S^3$ at each $y$ and $r$, with the boundary conditions $\theta \to \pi/2$ as $y^2 + r^2 \to \infty$, and possibly ending at some hypersurface where $\theta(y,r) = 0$.  

By symmetry, one extremal surface in this class will be $\theta = \pi/2$.   But we do not expect this to be the minimal surface.  Recall that in the previous section we found that the minimal boundary-anchored surface in AdS$_5 \times S^5$ for a boundary condition which cuts the $S^5$ in half is not the surface $\theta'= \pi/2$ which cuts the $S^5$ in half everywhere. We argued that it is instead a surface which pinches off to $\theta' =0$ near the boundary of the AdS factor.   Now recall that in the interior of the LLM geometry the $S^3$ of the asymptotic AdS$_5$  exchanges roles with an $\tilde{S}^3$ in the asymptotic $S^5$. Since the surface $\theta = \pi/2$ bisects the $S^3$ along the equator, in the IR AdS region it bisects the $S^5$ factor.  Given our reasoning about surfaces that bisect the $S^5$ in an AdS$_5 \times S^5$ geometry, we should expect that we can we can lower the area in the annulus geometry by allowing our candidate minimal surface to slip off the $S^5$ before it reaches the deep interior of the IR region.

In practice we need to be more careful; deforming the surface will increase the area in the UV region (which dominates in volume), so we must take care that this does not overwhelm the reduction from capping the surface off.  To argue this we make use of the intermediate metric, which has a domain of validity partially overlapping the domains of validity of the UV AdS and IR metrics.  We would expect the capping off to happen in this ``neck'' region, where we cross over from the UV AdS where $\theta=\pi/2$ is minimal, to the IR AdS where it is preferable to cap off. We will show that there are surfaces which cap off in the intermediate regime, for which the deviation from $\theta = \pi/2$ in the UV region is sufficiently small that the contribution of the UV region to the change in area is parametrically smaller than the decrease in area coming from the intermediate regime.

The geometry in the intermediate regime preserves the same $SO(6) \times SO(4)$ symmetry as in the IR regime. It is useful to make this symmetry manifest by making the coordinate transformation \eqref{s5coord}, so that the metric becomes
\begin{equation}
ds^2_{t=0} \approx \epsilon \sqrt{1+ e^{-4 \zeta}} e^{2\zeta} ( d\zeta^2 + d\theta^{'2} + \sin^2 \theta' d\Omega_4^2) + R (1 + e^{-4 \zeta})^{-1/2} d\tilde \Omega_3^2. \label{eq:intmettwo}
\end{equation}
In this coordinate transformation, $\theta = \pi/2$ maps to $\theta' = \pi/2$.  (The coordinate $\beta$ in \eqref{s5coord} becomes part of the $S^4$ metric $d\Omega_4^2$ in the above.)   In the full LLM solution, we would expect the minimal surface to involve a function of two variables, $\theta(r,y)$, which corresponds in these coordinates to taking  $\theta'(\zeta,\beta)$.  However, the enhanced symmetry in the intermediate regime suggests that we can find a minimal surface by taking $\theta' = \theta'(\zeta)$, as in our previous analysis of the IR regime. We will see below that the corrections from the UV regime for the surfaces we consider are small, so this should be a good approximation to the actual minimal surface. 

For a surface $\theta'(\zeta)$, the area functional is 
\begin{equation}
A = \sqrt{\epsilon^5 R^3} V_{S^4} V_{S^3} \int d \zeta e^{5 \zeta} \sqrt{1 + e^{-4 \zeta}} \sqrt{1 + (\partial_\zeta \theta')^2} \sin^4 \theta' .\label{eq:intarea}
\end{equation}
The resulting equation for the surface is 
\begin{equation} \label{zetaeq}
\partial_{\zeta}^2 \theta'  - (1+ (\partial_{\zeta} \theta')^2 ) (-\frac{3+5 e^{4 \zeta}}{1 + e^{4 \zeta}}  \partial_\zeta \theta' + 4 \cot \theta') = 0.
\end{equation}

\subsection{UV contributions}

We expect the minimal surface to approach $\theta' = \pi/2$ at large $\zeta$, where we patch on to the UV region. We therefore first consider a linearized analysis in this regime. The solution to the linearized version of \eqref{zetaeq}\ has the form $\theta' - \pi/2 \sim - \delta_0 e^{-a \zeta}$, with $a = 4$ or $a = 1$. These are fast and slow fall-off branches, analogous to the familiar normalizable and non-normalizable branches for a mode in AdS. We can expand \eqref{eq:intarea}\ to quadratic order in $\delta_0$ to approximate the gain or loss in area, and compare to the contribution in the UV region. 

If the surface approaches $\theta'=\pi/2$ at large $\zeta$ with a non-zero coefficient for the $a = 1$ solution, the integral over $\zeta$ is dominated by large values of $\zeta$, and the change in area from the $\theta' = \frac{\pi}{2}$ solution is of the order:
\be
	\Delta A_{int}^{a = 1} \sim - \sqrt{\eps^5 R^3} \delta_0^2 e^{3\zeta_{max}}.
\ee
While the solution appears to {\it lower}\ the area, we must also take the contribution from the UV region into account.
If we take $y,r \sim \eps^{\delta} R^{1-\delta}$ with $0 < \delta < \half$, the dominant contribution comes from a region in which the UV metric is a very good approximation. The angular deviation in the matching region is 
$\delta_0 e^{-\zeta_{max}} \sim \delta_0 \left(\frac{\eps}{R}\right)^{\half - \delta}$.  The UV contribution will have the form
\be
	\Delta A_{uv} \sim c_{uv} R^4 \delta_0^2 \left(\frac{\eps}{R}\right)^{1-2\delta}
\ee
Here $c_{uv}$ is a constant which will reflect the fact that the matching region $y,r \sim \eps^{\delta} R^{1-\delta}$ corresponds to the range ${\tilde \theta} - \frac{\pi}{2} \sim \left(\frac{\eps}{R}\right)^{\delta}$ in the UV metric. This covers a volume fraction of the UV $S^5$ of order $\left(\frac{\eps}{R}\right)^{2\delta}$, from the restricted range of ${\tilde \theta}$ and the smallness of the ${\tilde\phi}$ direction. If $c_{uv}$ scales with this volume fraction, then
\be\label{uva1}
	\Delta A_{uv}^{a=1} \sim c' \eps R^3 \delta_0^2 \gg |\Delta A_{int}|
\ee
If $c'$ is negative, then we arrive at a contradiction. We can simply cap the surface off at $\zeta \sim 1$, with a gain $(\eps^5 R^3)^{1/2} \ll \eps R^3$ in area, so that the area remains negative.  If we then deform the metric to the pure $AdS$ solution, the contribution of this cap remains small, and we have a surface in vacuum AdS with area less than the $\theta = \pi/2$ surface, contradicting the discussion in \S3.1.  Thus, if a solution exists with the $a = 1$ behavior in the intermediate regime, it is not a minimal surface.

Instead, we will consider surfaces which approach $\theta'=\pi/2$ at large $\zeta$ with the fast fall-off, that is $a = 4$. In this case, $\theta' = \frac{\pi}{2} - \delta_0 e^{-4\zeta}$, and the contribution from large $\zeta$ in \eqref{eq:intarea} is suppressed. At the matching point $e^{\zeta_{max}} \sim \sqrt{\frac{R}{\eps}}$,  $\theta - \pi/2 \sim \delta_0 \frac{\eps^2}{R^2}$, and the contribution to the area from the UV region scales as 
\be
	\Delta A^{a=4}_{uv} = d_{uv} \eps^4 \delta_0^2
\ee
where $d_{uv}$ is a positive constant of order $1$. This is smaller than the contribution $\delta_0^2 \sqrt{\eps^5 R^3} e^{-3\zeta_{ir}}$ from the intermediate region, where $\zeta_{ir}$ is the scale where the linearized approximation breaks down (we will find this happens while the metric is still well approximated by \eqref{eq:intmettwo}). Thus, for the $a = 4$ solutions, the contribution from the UV region is negligible, and we can consistently calculate the change in area in the intermediate region. 

One additional caveat, mentioned above, is that we assumed more symmetry in the intermediate region than we expect the exact solution to have.   This allowed us  to write $\theta^\prime$ as a function of a single variable $\zeta$.  In general,  due to the boundary matching that we must do  at large $\zeta$, the solution will have the form $\theta'(\zeta,\beta)$ (where $\beta$ is a coordinate in the $S^4$ with metric $d\Omega_4^2$ in \eqref{eq:intmettwo}).  However, we expect that this symmetry breaking will {\it increase}\ the area in the intermediate regime.  Thus, since the dominant change in area occurs in the intermediate regime, and the deviation from $\theta - \pi/2$ is small at the transition point to the UV metric, the symmetry-breaking component of the true minimal surface will be suppressed.  In other words, if we choose the surface in the intermediate region to only depend on $\zeta$ we get a smaller area, and thus although some $\beta$ dependence will be induced by the matching with the UV, it will be suppressed since it is advantageous for the minimal surface to depend to be a function only, or mostly, of $\zeta$. 

\subsection{Minimal surface}

\begin{figure}[t!]
\centering 
\begin{overpic}[width=16cm, trim={0cm 0cm 0cm 0cm},clip]{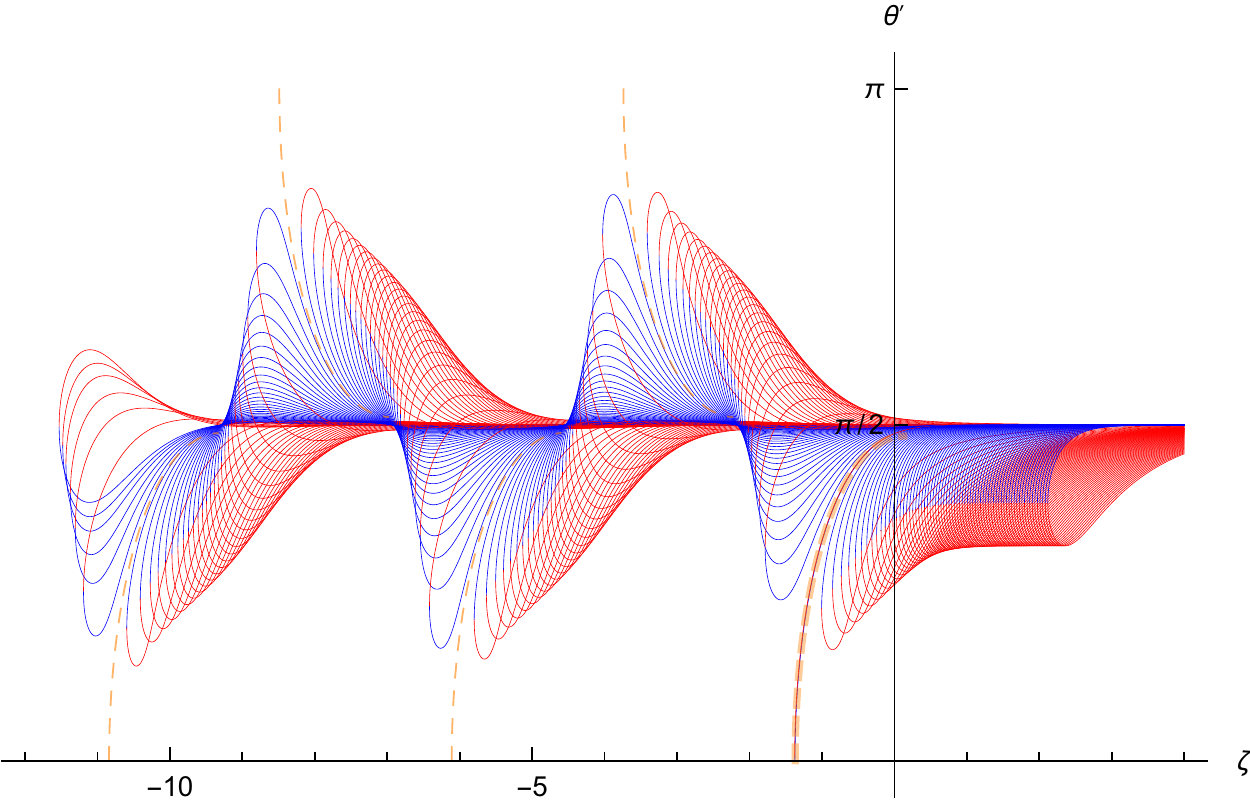}
\put (51,60) {\large{$\twoheadleftarrow$ IR AdS}}
\put (75,60) {\large{UV AdS $\twoheadrightarrow$} }
\end{overpic}
\caption{Intermediate regime numerical solutions with varying values of $\theta'(\zeta_{max})$, taking $\zeta_{max} = 4$. The surfaces plotted are the part of the subset that approach the $a=4$ linearised solution, such that they are candidate minimal surfaces. The blue section of a curve shows where the surface move inwards from $\zeta_{max}$, the red section where it moves outward. Equivalent flipped solutions $\theta' \to \pi - \theta'$ are not shown for clarity. Orange dashed lines are truncated series solutions of \eqref{zetaeq}, showing the discrete values of $\theta'(\zeta_{max})$ for which the surface reaches $\theta' = 0$. The thickest orange line hitting $\theta' = 0$ at $\zeta_{min} = - 1.37$, through which one numerical solution passes, shows the true minimal surface which approaches $\theta' = \pi/2$ at large $\zeta$ in this LLM geometry.} \label{isplot}
\end{figure}

We can find extremal surfaces by solving the equation \eqref{zetaeq} numerically, with the boundary condition that we approach the linearized solution with $a=4$ at large $\zeta$. We solve for $\delta(\zeta) = \theta'-\pi/2$, taking the boundary condition $(1/\delta) (d \delta / d\zeta) = -4$
at some large radius $\zeta_{max}$, and shoot in. The solutions are shown in figure \ref{isplot}. 

We find that the solutions have an interesting structure: the solution for generic values of $\delta(\zeta_{max})$ encounters a turning point where $\partial_\zeta \theta' \to \infty$, and then an extremum where $\partial_\zeta \theta' =0$, and then returns to large $\zeta$ with $\theta' \to \pi/2$. These generic surfaces do not satisfy our boundary conditions, as they would intersect the boundary twice. 

Those surfaces whose deviation from $\theta' = \pi/2$ at  $\zeta_{max}$ is sufficiently small that they reach the IR region begin to oscillate around $\theta' = \pi/2$ as described by \eqref{eq:IR_oscillations}. The surface's deviation from $\pi/2$ grows exponentially until the linearized solution is invalid. The candidate minimal surfaces arise for discrete values of the initial conditions, where the first turning point lies at $\theta^\prime = 0$ or $\theta^\prime = \pi$, and the surface smoothly caps off. If the solution hits $\theta' = 0$ at $\zeta = \zeta_0$, smoothness requires  that $\zeta - \zeta_0 \sim (\theta')^2 + \ldots$. Indeed, we can expand \eqref{zetaeq}\ about $\theta' = 0, \zeta = \zeta_0$ to find $\theta'(\zeta - \zeta_0) \sim (\zeta - \zeta_0)^{1/2}$. Of the solutions which cap off, the solution with the lowest area caps off at the largest value of $\zeta$, $\zeta =  -1.37$.  

The minimal surface has an area $1.43 \sqrt{\epsilon^5 R^3}$ less than the $\theta = \pi/2$ surface. As expected from the general scaling argument, the reduction in area scales as $\sqrt{\eps^5 R^3}$. Thus, from the analysis in the intermediate regime, which is reliable for the surfaces which approach $\theta=\pi/2$ in the UV on the fast fall-off $a=4$ branch, we learn that there is a surface in the LLM annulus geometry which bisects the $S^3$ of the AdS$_5$ factor at infinity at $\theta = \pi/2$ which has smaller area than the surface that remains at  $\theta = \pi/2$ throughout. The surface caps off in the intermediate regime, at the edge of the region where the IR AdS metric begins to be a good approximation.  Thus, the minimal surface barely reaches the interior IR AdS regime.

\section{Discussion}
\label{disc}

We have found new examples of entanglement shadows in LLM geometries. We analysed a specific example where the geometry is simple enough that we could approximately determine the location of the minimal surface, but we expect this behavior to be more general. The essential reason for the change in the minimal surface is that the $S^3$ that our minimal surface divides goes from having a volume which decreases as we moved inwards through the UV region, to being essentially constant as we enter the IR region. When the volume of the sphere is decreasing, the minimal area surface stretches across the ball, as in flat space. But when the volume of the sphere becomes constant, the minimal surface wraps around the sphere at nearly constant radius, as on a cylinder. Thus we would expect that such shadows would be seen in any LLM geometry where the volume of the $S^3$ becomes approximately constant in the interior. That is, for cases where there are one or more white regions inside a black region.

Our story is not, however, completely generic. If we consider instead an LLM geometry with two black discs, when the discs are well-separated we can treat the region near each disc as an approximate copy of AdS$_5 \times S^5$, and we would not expect there to be a shadow region. There is also no reason to expect a shadow for small fluctuations in the shape of the AdS black disc geometry.

As we vary $\theta_0$, the location of the minimal surface jumps at $\theta_0 = \pi/2$ from passing above the shadow region to passing below it. However, there is an extremal {\it non}-minimal surface at $\theta = \pi/2$ which passes through the shadow region. Each point in the shadow region lies on such a surface for some choice of division of the boundary. Therefore it would be very interesting if this non-minimal surface could be interpreted in terms of a CFT observable, similar to the entwinement of \cite{Balasubramanian:2014sra,Balasubramanian:2016xho}. One of the advantages of conducting the analysis in the LLM context is that the dual CFT states are known precisely, so we can explore the entanglement structure of these states and see if they lead to interesting observables. The states correspond to Young tableaux, and it is intriguing to speculate that the $S_N$ structure encoded in these tableaux could play a role here, as the $S_N$ symmetry of the symmetric orbifold did in the entwinement story. 

Another possibly related discussion is that in \cite{Karch:2014pma}, which studies surfaces in a Coulomb branch geometry in which $SU(N)$ is broken to $SU(m)\times SU(N-m)$ where $m, N$ are of similar order.   They construct surfaces that bisect the $S^5$ factor at large radius (scales above the symmetry breaking scale) and pass between the two IR AdS factors; whether the minimal surface in this class enters one region, the other, or neither, depends on where one places the cutoff. They conjecture that surfaces passing between the two factors measure the entanglement between the light fields associated with each unbroken gauge factor.

As the cutoff is removed, all of these surfaces bisect the $S^5$ at the equator, as proven in \cite{Graham:2014iya}. However, following our discussion above, none of these are minimal for any cutoff; there are always arbitrarily small extremal surfaces which only extend inward for a small distance away from the cutoff surface. In Figure 3 of \cite{Karch:2014pma}, these would be surfaces that pass $r = 0$ at larger values of $y$ than are shown. Nonetheless, the surfaces studied in \cite{Karch:2014pma} are extremal, and as with the entwinement story there may be a plausible inerpretation in terms of entanglement explicitly involving the matrix degrees of freedom of the theory. 


\section*{Acknowledgements}

We are grateful for useful discussions with Aitor Lewkowycz, Onkar Parrikar and Charles Rabideau.  AL and AR are supported in part by DOE grant DE-SC0009987.    SFR is supported in part by STFC under consolidated grant ST/L000407/1.  VB was supported  by the Simons Foundation (\#385592, It From Qubit Collaboration) and by DOE grant DE-FG02-05ER- 41367.


\bibliographystyle{JHEP}
\bibliography{bubbles}

\end{document}